\documentstyle[12pt]{article}

\begin{document}

\begin{titlepage}

\begin{flushright}
{\sc FERMILAB-PUB-93/307-T}\\
{\sc CMU-HEP93-25; DOE-ER/40682-50}\\
{\sc TRI-PP-93-103}\\ [.2in]
{\sc \today}\\ [.5in]
\end{flushright}

\begin{center}
{\LARGE Direct CP violation in $b \rightarrow d J/\psi$ decays}\\ [.5in]
{\large Isard Dunietz}\\[.1in]
{\small Fermilab, P.O.\ Box 500, Batavia, IL 60510}\\ [.2in]
{\large and}\\ [.2in]
{\large Jo\~{a}o M. Soares}\\ [.1in]
{\small Department of Physics, Carnegie Mellon
University, Pittsburgh, PA 15213\\ [.1in]
and\\ [.1in]
TRIUMF, 4004 Wesbrook Mall, Vancouver, BC Canada V6T 2A3}\\ [.5in]

{\normalsize\bf Abstract}\\ [.2in]
\end{center}
{\small
We investigate the possibility of observing direct CP violation in
self-tagging B-meson decays of the type $b \rightarrow d J/\psi$.
The CP asymmetry can be generated due to strong or electromagnetic
scattering in the final state, or due to long distance effects.
The first two contributions give asymmetries of $a \, few \,\times 10^{-3}$,
in the standard model. The long distance effects are hard to estimate, but
it cannot be excluded that they yield asymmetries of about $1\%$.\\
PACS: 13.25.+m, 11.30.Er, 12.15.Ff.}

\end{titlepage}

\section*{}

The standard model (SM) predicts the existence of relative CP-odd phases
between different B-meson decay amplitudes (direct CP violation).
If two such amplitudes contribute coherently to a decay $B \rightarrow f$,
they may generate a CP asymmetry
\begin{equation}
a_{CP} = \frac{\Gamma(B \rightarrow f) -
\Gamma(\bar{B} \rightarrow \bar{f})}{\Gamma(B \rightarrow f) +
\Gamma(\bar{B} \rightarrow \bar{f})}.
\label{eq:1}
\end{equation}
This type of asymmetry does not require mixing, and it can occur in
self-tagging decay modes. These are decays such that
$B \not\rightarrow \bar{f}$ and $\bar{B} \not\rightarrow f$
(e.g. decays of charged B-mesons), and so the distinction between
$B$ and $\bar{B}$ is immediate, given the final state. As a result,
the experimental sensitivity is improved by one order of magnitude
with respect to the case of the neutral B-meson decays into CP eigenstates,
where the tagging is more involved \cite{snowmass:tagging}.

In the SM, the numerator of eq.~\ref{eq:1} is proportional to
$\sin^6 \theta_C$, and so the asymmetry can be significant only
for decays that are strongly Cabibbo suppressed. The asymmetries
in rare decays to charmless states (with neither bare nor hidden charm)
have been calculated in earlier works
\cite{bss:mechanism,charmless:asym,wolf:cpt}:
for typical values of the parameters in the Cabibbo-Kobayashi-Maskawa (CKM)
matrix, the 1-loop level decays such as $b \rightarrow d s \bar{s}$ and
$b \rightarrow d \gamma$ have asymmetries of about $5\%$. The tree level
decays, such as $b \rightarrow d u \bar{u}$, and the Cabibbo favored
$b \rightarrow s$ transitions have asymmetries that are one order of
magnitude lower. Exclusive decays have also been discussed: their asymmetries
should be of the same order as those in the corresponding semi-inclusive
processes, but (except for the case of the radiative decays) there are
uncertainties from the hadronic form factors, as well as non-perturbative
contributions that are difficult to estimate and could be large.
The branching ratios are of the order of $10^{-6}$ for decays such as
$B^- \rightarrow K_S K^-$ and $B_s^0 \rightarrow \bar{K}^{\ast 0} \gamma$.
Only future experiments, with appropriate triggering and good $K/\pi$
separation, will be able to probe the asymmetries in these modes, at a
level anywhere near the SM predictions \cite{snowmass:delta}.

The self-tagging decays of the type $b \rightarrow d c \bar{c}$,
where the charm-anticharm pair forms a $J/\psi$, may show CP violating
effects \cite{isi:ew}. The size of those effects is investigated in this paper.
The modes with a $J/\psi$ are particularly attractive from the experimental
point of view, as one can trigger on the $J/\psi$ {\it via} its dilepton
decay mode.
Moreover, this is feasible at hadronic accelerators (as demonstrated
by the current results from CDF \cite{CDF:Bphysics}), where large numbers
of B-mesons can be produced. The branching ratio for the
$b \rightarrow d J/\psi$ transition is approximately $\sin^2 \theta_C
\times BR(B \rightarrow J/\psi \; anything) \simeq  5 \times 10^{-4}$,
and the branching ratios for the exclusive decays, such as
$B^- \rightarrow J/\psi \pi^-$ or $B_s^0 \rightarrow J/\psi \bar{K}^{\ast 0}$,
are one order of magnitude lower. With an expected sample of $10^{10}$
B-mesons, and provided a detection efficiency (including $K/\pi$ separation)
no lower than $10\%$, the asymmetries in the exclusive modes can be probed
down to around $1\%$ (at the $3\sigma$ level).

Here, we estimate the value of the CP asymmetry that is predicted by the
SM for these modes. As is always the case for the asymmetries that are due
to direct CP violation, the difficulty lies in estimating the strength of
the final state interactions. We will point out that the 1-gluon mediated
scattering \cite{bss:mechanism}, that is expected to dominate in the case
of the charmless decays that were mentioned above, does not contribute to
the asymmetry in $b \rightarrow d J/\psi $. We then proceed to discuss
the main contributions to the final state scattering in this case. For
each one of them, the resulting asymmetry is either calculated, or an
attempt is made to give a reasonable estimate. We will find that the
asymmetry that is predicted in the SM lies at the level, or slightly
below, the experimental sensitivity that is expected in the near future.

\section*{}

The tree contribution to the $b \rightarrow d J/\psi$ decay amplitude is
$V_{cb} V_{cd}^\ast T_{d\psi}$. With
\begin{equation}
< 0 | \bar{c} \gamma_\mu c | J/\psi > = m_{J/\psi} f_{J/\psi} \epsilon_\mu ,
\label{eq:2a}
\end{equation}
factorization gives
\begin{equation}
T_{d\psi} =  G_F \frac{1}{\sqrt{2}}
a_1 m_{J/\psi} f_{J/\psi} \epsilon_\mu^\ast \bar{u}_d \gamma^\mu
(1-\gamma_5) u_b .
\label{eq:3}
\end{equation}
Because the $c\bar{c}$ pair must be in a color singlet, there is a
color suppression factor $a_1$. Including the leading-logarithm
QCD corrections, $a_1 = C_1(m_b) + C_2(m_b)/N_c$ \cite{colorsuppr:disc}
and the Wilson coefficients are \cite{wilsoncoeff:Buras}
\begin{eqnarray}
C_1(m_b) &=& 0.25 \nonumber \\
C_2(m_b) &=& - 1.11 ,
\label{eq:4}
\end{eqnarray}
for $\Lambda^{(4)}_{\overline{MS}} = 200 MeV$ \cite{lambda:MS}
and $m_b = 4.8 GeV$.
The CP asymmetry arises from the interference
between the tree decay amplitude, and any additional term
that has a different CP-odd phase and a different CP-even phase.
The latter phase appears when the additional contribution is due to the
decay into an on-mass-shell intermediate state that then scatters
into $d J/\psi$, through final state interactions. The intermediate state
that is favored is $du\bar{u}$, since it is fed by a tree level decay amplitude
$V_{ub} V_{ud}^\ast T_{du\bar{u}}$. When the scattering $u\bar{u} \rightarrow
J/\psi$ can be treated perturbatively, we have
\begin{equation}
\bar{A} \equiv A(b \rightarrow d J/\psi) = V_{cb}V_{cd}^\ast  T_{d\psi}
+ i \frac{1}{2} V_{ub}V_{ud}^\ast  T_{du\bar{u}} A(u\bar{u} \rightarrow
J/\psi).
\label{eq:6}
\end{equation}
The interference between the two terms on the RHS gives the
difference
\begin{equation}
|A|^2 -|\bar{A}|^2
= 2 Im\{V_{cb}^\ast V_{cd} V_{ub} V_{ud}^\ast\} T_{d\psi}^{\ast}
T_{du\bar{u}} A(u\bar{u} \rightarrow J/\psi),
\label{eq:7}
\end{equation}
(the sum over the spin, color and phase space of the intermediate state
is implied), and hence the CP asymmetry of eq.~\ref{eq:1}.

In the rare decays into charmless final states that have been studied in the
literature, the final state scattering occurs through order $\alpha_s$
diagrams \cite{bss:mechanism}. For example, for the asymmetry in
$b \rightarrow ds\bar{s}$, the additional term in the amplitude is mostly due
to $b \rightarrow du\bar{u} \stackrel{g}{\rightarrow} ds\bar{s}$.
In the case of $b \rightarrow du\bar{u} \rightarrow d J/\psi$ (and
within the factorization approximation),
the 1-gluon scattering cannot contribute as the $J/\psi$ is a color singlet.
Because the strong interaction is invariant under charge conjugation,
the 2-gluon amplitude also vanishes. The QCD scattering can then occur only
at the order $\alpha_s^3$, and it is comparable to the electromagnetic
scattering \cite{charmonium:Trottier}.

A rough estimate shows that the QCD and QED scatterings from the $du\bar{u}$
intermediate state are indeed of similar strength. The ratio of these two
contributions,
\begin{equation}
A(u\bar{u} \stackrel{QCD}{\rightarrow} J/\psi)
/A(u\bar{u} \stackrel{QED}{\rightarrow} J/\psi),
\label{eq:8}
\end{equation}
is of the same order as $|\Gamma(J/\psi \stackrel{QCD}{\rightarrow} u\bar{u})/
\Gamma(J/\psi \stackrel{QED}{\rightarrow} u\bar{u})|^{1/2}$.
The QED width is equal to $2/3$ of $\Gamma(J/\psi \stackrel{QED}{\rightarrow}
hadrons)=(17.0 \pm 2.0)\% \times \Gamma_{J/\psi}$. The QCD width is some
fraction (not too different from the $1/3$ of a naive constituent picture)
of $\Gamma(J/\psi \stackrel{QCD}{\rightarrow} hadrons) = \Gamma(J/\psi
\rightarrow hadrons) - \Gamma(J/\psi \stackrel{QED}{\rightarrow} hadrons) =
(69.0 \pm 2.8)\% \times \Gamma_{J/\psi}$. It follows that the ratio in
eq.~\ref{eq:8} is $\sim 1.5$.

Of the two similar contributions, the electromagnetic one is simpler to
calculate. The convolution of the amplitude for the decay
$b \rightarrow du\bar{u}$ and that for the 1-photon scattering
$u\bar{u} \rightarrow J/\psi$ gives
\begin{eqnarray}
T_{du\bar{u}} A(u\bar{u} \rightarrow J/\psi)
&=& - G_F \sqrt{2} \alpha Q_u Q_c a_1
\nonumber \\
& & m_{J/\psi} f_{J/\psi} \epsilon_\mu^\ast \bar{u}_d \gamma^\mu
(1-\gamma_5) u_b \nonumber \\
&=& - T_{d\psi} 2 \alpha  Q_u Q_c
\label{eq:9}
\end{eqnarray}
The contribution to the asymmetry in the semi-inclusive decay
$b \rightarrow d J/\psi$ is then
\begin{equation}
a_{CP} \simeq - \eta \alpha 8/9 = - 0.3\%,
\label{eq:10}
\end{equation}
where $\alpha(m_{J/\psi}) = 1/133$, and $\eta = 0.4$ has been chosen as a
typical value for the CKM parameter, within the present bounds
\cite{CKM:update}. This result should also hold for the exclusive decays.
The reason is that the two terms in the decay amplitude (analogous to those
in eqs.~\ref{eq:3} and \ref{eq:9}) have the same operator structure.
Then the hadronic matrix element can be factored out, and the expression for
the asymmetry is that given in eq.~\ref{eq:10}.

So far we have ignored the effect of the intermediate state $dc\bar{c}$.
For the case of the inclusive decay, $b \rightarrow dc\bar{c}$, that effect
is just a re-scattering of the final state. It does not generate two
amplitudes (with different CKM phases) that can interfere,
and so there is no contribution to the asymmetry \cite{wolf:cpt}.
But for the exclusive or semi-inclusive cases that we are discussing,
the situation is different. It has been pointed out by Wolfenstein
\cite{wolf:cpt} that contributions to the asymmetry, from intermediate
states with the same quark content as the final state, will arise, once
the small penguin amplitudes are added to the tree amplitudes considered so
far.
For example, the amplitude for a decay such as $B^- \rightarrow J/\psi \pi^-$
becomes
\begin{eqnarray}
A(B^- \rightarrow J/\psi \pi^-) &=& V_{cb} V_{cd}^\ast T_{\psi\pi^-}
+ V_{tb} V_{td}^\ast P_{\psi\pi^-} \nonumber \\
&+& i \frac{1}{2} \sum_X
\left( V_{cb} V_{cd}^\ast T_{X}
+ V_{tb} V_{td}^\ast P_{X} \right) A(X \rightarrow J/\psi \pi^-).
\label{eq:11}
\end{eqnarray}
The penguin amplitudes are the terms proportional to
$V_{tb} V_{td}^\ast$, and we have included the absorptive part
due to the intermediate states $X$. These are the states
$D^0 D^-$, $D^{\ast -} D^0$, $J/\psi \rho^-$, etc., that have the same
quark content as the final state $J/\psi \pi^-$ (for clarity, we now
omit the absorptive part due to
$b \rightarrow du\bar{u} \rightarrow dc\bar{c}$ that was discussed before).
Because the matrix elements of the tree and penguin operators depend on
the hadronic states, the penguin/tree ratios
$P_{\psi\pi^-}/T_{\psi\pi^-}$ and $P_{X}/T_{X}$ will in general be different.
Then, the dispersive and absorptive parts of the amplitude in eq.~\ref{eq:11}
will have different CKM phases, and so the states $X$ will contribute to
the CP asymmetry with
\begin{equation}
a_{CP} \simeq Im\{\frac{V_{tb} V_{td}^\ast}{V_{cb} V_{cd}^\ast}\}
 \sum_X \frac{T^{\ast}_{\psi\pi^-} T_X A(X \rightarrow J/\psi \pi^-)}
{|T_{\psi\pi^-}|^2} (\frac{P_X}{T_X} - \frac{P_{\psi\pi^-}}{T_{\psi\pi^-}}).
\label{eq:12}
\end{equation}
The final state scatterings $A(X \rightarrow J/\psi \pi^-)$ are
long distance effects that are hard to estimate. We will compute the
asymmetry due to some of the intermediate states $X$, leaving the ratio
\begin{equation}
 \xi_X  \equiv \frac{T^{\ast}_{\psi\pi^-} T_X
A(X \rightarrow J/\psi \pi^-)}  {|T_{\psi\pi^-}|^2}
\label{eq:13}
\end{equation}
as an undetermined parameter. In particular, we will look at intermediate
states such as $D^0 D^-$, where $c\bar{c}$ is not required to form a color
singlet. There, the amplitude for the decay $B \rightarrow X$ is not color
suppressed, and the parameter $\xi_X$ may be larger. Notice that, if the
branching ratio for $B^- \rightarrow J/\psi \pi^-$ can be measured with
sufficient precision (and if the short distance contribution is well
understood), then some information can be obtained on the strength of the
final state scatterings (barring possible cancellations
between the different intermediate states $X$). For the moment, let us just
assume that $A(X \rightarrow J/\psi \pi^-)$  can be treated perturbatively
(so that eqs.~\ref{eq:11} and \ref{eq:12} remain valid).

The tree and penguin decay amplitudes are calculated from the
effective Hamiltonian
\begin{eqnarray}
 H_{eff} &=& - \frac{G_{F}}{\sqrt{2}}  \,[ V_{ub} V_{ud}^\ast
( C_{1} {\cal Q}_{1}^u +
C_{2} {\cal Q}_{2}^u ) + V_{cb} V_{cd}^\ast
( C_{1} {\cal Q}_{1}^c +
C_{2} {\cal Q}_{2}^c ) \nonumber \\
& & + V_{tb} V_{td}^\ast \sum_{k=3}^{6} C_{k} {\cal Q}_{k}  + h. c. ]
 \, ,    \label{eq:14}
\end{eqnarray}
where
\begin{eqnarray}
{\cal Q}_{1}^l &=& \bar{d} \gamma^{\mu} (1 - \gamma_5)  b \:\,
\bar{l} \gamma_{\mu}  (1 - \gamma_5) l  \nonumber \\
{\cal Q}_{2}^l &=&  \bar{l} \gamma^{\mu}  (1 - \gamma_5) b \:\,
\bar{d} \gamma_{\mu}  (1 - \gamma_5) l  \nonumber \\
{\cal Q}_{3} &=& \sum_{l=u,d,s,c,b} \bar{d}\gamma^{\mu}  (1 - \gamma_5) b \:\,
\bar{l}\gamma_{\mu}  (1 - \gamma_5) l  \nonumber \\
{\cal Q}_{4} &=& \sum_{l=u,d,s,c,b} \bar{l}\gamma^{\mu}  (1 - \gamma_5) b \:\,
\bar{d}\gamma_{\mu}  (1 - \gamma_5) l  \nonumber \\
{\cal Q}_{5} &=& \sum_{l=u,d,s,c,b}  \bar{d}\gamma^{\mu}   (1 - \gamma_5) b
\:\,
\bar{l}\gamma_{\mu}   (1 + \gamma_5) l  \nonumber \\
{\cal Q}_{6} &=&  - 2 \sum_{l=u,d,s,c,b} \bar{l}   (1 - \gamma_5) b \:\,
\bar{d}  (1 + \gamma_5) l     ,   \label{eq:15}
\end{eqnarray}
and, for $\Lambda_{\bar{MS}}^{(4)} \simeq 200 MeV$, the Wilson coefficients are
\cite{wilsoncoeff:Buras}
\begin{eqnarray}
  C_3 (m_{b})&=&0.011 \nonumber \\
  C_4 (m_{b})&=&-0.026 \nonumber \\
  C_5 (m_{b})&=&0.008 \nonumber \\
  C_6 (m_{b})&=&-0.032    \label{eq:15a}
\end{eqnarray}
($C_1$ and $C_2$ were given in eq.~\ref{eq:4}). We use factorization and
neglect the terms of order $1/N_c$ \cite{colorsuppr:disc}.
For the decays of the type $b \rightarrow d J/\psi$,
the penguin to tree ratio is
\begin{equation}
 \frac{P_{d\psi}}{T_{d\psi}} = \frac{C_3+C_5}{C_1} = 0.076 .
\label{eq:16}
\end{equation}
Whereas for $B^- \rightarrow D^0 D^-$, and for some other color favored
decays, we find
\begin{eqnarray}
 \frac{P_{D^-D^0}}{T_{D^-D^0}} &=&
\frac{1}{C_2} (C_4+2 C_6\frac{1}{m_b-m_c} \frac{m_D^2}{m_c+m_d})  = 0.064
\nonumber\\
 \frac{P_{D^-D^{\ast 0}}}{T_{D^-D^{\ast  0}}} &=&
\frac{1}{C_2} (C_4-2 C_6\frac{1}{m_b+m_c} \frac{m_D^2}{m_c+m_d})  = 0.0021
\nonumber\\
 \frac{P_{D^{\ast -}D^0}}{T_{D^{\ast -}D^0}} &=&
 \frac{P_{D^{\ast -}D^{\ast 0}}}{T_{D^{\ast -}D^{\ast 0}}} =
\frac{C_4}{C_2}  = 0.023
\label{eq:17}
\end{eqnarray}
(with $m_c=1.5 GeV$ and $m_d \ll m_c$). The equations of motion
have been used to relate the different matrix elements, so that the hadronic
uncertainties always cancel in the penguin/tree ratios.
For some of these ratios, the effect of 1-loop electroweak corrections
\cite{isi:ew} can be significant. A thorough analysis of such contributions,
including QCD corrections, can be found in ref.~\cite{buras:ew}. Using the
results in there, we derive the corrected values for the penguin/tree ratios:
$P_{d\psi}/T_{d\psi} = 0.042$ and
$P_{D^-D^{\ast 0}}/T_{D^-D^{\ast  0}} = 0.0012$;
whereas for the other decays, the electroweak effects are not larger
than $10\%$. Replacing these values in eq.~\ref{eq:12}, one finds
contributions to the asymmetry of about
\begin{equation}
 a_{CP} = \xi_X \times 1\%
\label{eq:18}
\end{equation}
(for $\eta = 0.4$).
This number should give us a rough idea of the size of the asymmetries
(for either the semi-inclusive or the exclusive cases), that are expected
from the long distance effects. Although, there are contributions from
many channels that add with different signs, it is unlikely that large
cancellations or enhancements will occur. Therefore, according to the size
of $\xi_X$, the contribution in eq.~\ref{eq:18} could be comparable to the
short distance effects described before, and give an asymmetry slightly below
the expected experimental sensitivity. But it could also be the dominant
effect, and then the asymmetry will be within reach of the ongoing
experiments at the Tevatron.

We should stress that our results were derived using
factorization, together with the prescription of dropping $1/N_c$
contributions to the hadronic matrix elements in the decay amplitudes
\cite{colorsuppr:disc}. This is the same prescription that is successful
in predicting the branching ratios for the decays of the type
$b \rightarrow s J/\psi$. Different results would follow,
for example, by taking $N_c=3$. In that case, some new mechanism must
contribute to the color suppressed decays, that would affect the branching
ratio, and most certainly, also the asymmetry.

\section*{Acknowledgements}

We wish to thank Lincoln Wolfenstein for
pointing out to us the importance of the long distance effects,
and Howard Trottier for discussions.
Part of this work was accomplished within the scope of the
Fermilab Summer Visitors Program; J. M. S. is
grateful to the Theory Group at Fermilab for their hospitality.
This work was partly supported by the U. S. Department of Energy
under Contracts Nos. DE-AC02-76CH03000 and DE-FG02-91ER40682, and by
the Natural Science and Engineering Research Council of Canada.
I. D. was supported in part by the Texas National
Research Laboratory Commission under Award No. FCFY9303.


\begin{thebibliography}{Abcd:efgh}
\bibitem{colorsuppr:disc}
The magnitude of the color suppression factor $a_1$ can be measured in
the Cabibbo favored decays, of the type $b \rightarrow s J/\psi$.
The branching ratios for the exclusive decays $B \rightarrow J/\psi K$
and $B \rightarrow J/\psi K^\ast$ give \cite{colorsuppr:exp}
$|a_1| = 0.23 \pm 0.01 \pm 0.01$,
and a very similar value follows from the branching ratios for the
semi-inclusive decays $B \rightarrow J/\psi \; anything$
and $B \rightarrow \psi^\prime \; anything$.
This is very different from the theoretical result
$|a_1(N_c=3)| = |C_1(m_b) + C_2(m_b)/3| = 0.12$.
As pointed out in ref.~\cite{Nc:infinity}, a much better agreement
with the experimental value is obtained from
the leading order in the $1/N_c$ expansion: $a_1(N_c=\infty) = C_1(m_b)$.
The success of this approximation has been justified, for some cases,
by the existence of other contributions of order $1/N_c$, that appear
beyond the factorization approximation, and thus were not included in
eq.~\ref{eq:3} \cite{Nc:factoriz}. The procedure of taking $N_c=\infty$
is not entirely satisfactory (in particular, it appears to give the wrong sign
for $a_1$ \cite{colorsuppr:exp}), but we will adopt it, for the lack of
a better model. It should be pointed out that this is only relevant
in the last part of this work, when estimating the penguin contributions
for which there is no experimental guidance so far.
\end{thebibliography}
\end{document}